\documentclass[aps,pra,twocolumn,groupedaddress]{revtex4-1}

\usepackage{amsfonts,amsmath,amsthm,amssymb,amsbsy,graphicx,bbold}

\begin{document}


\title{Process Characterisation with Monte-Carlo Wave-Functions}


\author{J. Gulliksen}

\author{D. D. Bhaktavatsala Rao}

\author{K. M\o lmer}
\affiliation{Department of Physics and Astronomy, Aarhus University, Ny Munkegade 120, DK-8000 Aarhus C, Denmark}
	

\date{\today}

\begin{abstract}
We present a numerically efficient method for the characterisation of a quantum process subject to dissipation and noise. The master equation evolution of a maximally entangled state of the quantum system and a non-evolving ancilla system is simulated by Monte-Carlo wave-functions. We show how each stochastic state vectors provides quantities that are readily combined into an average process $\chi$-matrix. Our method significantly reduces the computational complexity in comparison with standard characterisation methods. It also readily provides an upper bound on the trace distance between the ideal and simulated process based on the evolution of only a single wave function of the entangled system.
\end{abstract}


\maketitle

\section{Introduction}
Characterisation of quantum-dynamical systems is a prerequisite for high fidelity quantum computing and information protocols. Two tools created for this purpose are quantum-state and quantum-process tomography. Quantum-state tomography takes the measurement data of a quantum system's unknown state and identifies the representative density operator $\rho$. Quantum-process tomography is concerned with experimentally characterising a process $\mathcal{E}$ so that the output state may be predicted from any given input state, $\rho\to\mathcal{E}(\rho)$.

To completely characterise a N-qubit process, standard quantum process tomography (SQPT)~\cite{nielsen2010quantum,chaung97proctom,PhysRevLett.78.390} requires implementation on a number of input states that scales exponentially with the number of qubits. Avoiding the preparation of many different input states ancilla-assisted quantum process tomography (AAPT)~\cite{choiproof_leung,PhysRevLett.86.4195,PhysRevLett.90.193601, PhysRevLett.91.047902} was proposed as an alternative. Here, the information of all input states is encoded into a single maximally entangled system-ancilla state in a doubled Hilbert space. While both SQPT and AAPT rely on state tomography of the output state, direct characterisation of quantum dynamics~\cite{PhysRevLett.97.170501,PhysRevA.75.062331} directly addresses features of the underlying dynamics via suitable ``probe'' systems and corresponding measurements. For more details and an investigation of resource demands for each of these strategies see~\cite{PhysRevA.77.032322}.

If a quantum system is subject to unitary and dissipative dynamics that are completely known, it may still be a non-trivial task to determine the resulting process $\mathcal{E}$ from a general input state to corresponding output state. Solving that problem is highly relevant in quantum information science, as it quantifies the fidelity of gates for different input states and may eventually point to the optimal application of these gates in algorithms. This article will present a procedure to numerically determine the quantum process $\mathcal{E}$ from the master equation governing time-dependent system dynamics. Monte-Carlo wave-functions~\cite{carmichael1993open,PhysRevLett.68.580,Molmer:93,PhysRevA.45.4879} have allowed a numerically efficient alternative to the ordinary master equation approach to obtain the time dependent density matrix $\rho(t)$. Indeed, a Monte-Carlo simulation of a quantum system with a Hilbert space of size $D$ ($D\gg 1$) involves far less variables ($\sim D$) than its master equation counterpart ($\sim D^2$). In this paper we simulate the evolution of our system from an initial maximally entangled state with a non-evolving ancilla system, and we show that the Monte-Carlo wave-functions for this combined system can be directly processed to yield information about the quantum process $\mathcal{E}$.

The paper is organized as follows. In Sec. II, we introduce the system master equation as well as the general problem of characterizing the time evolution as a quantum process applied to any initial state. In Sec. III, we introduce part of the formalism needed for process characterisation and describe density matrix schemes that yield the  process matrix $\chi$. In Sec. IV, we briefly recall the Monte Carlo wave-function method and relevant aspects of its implementation in this work. In Sec. V, we show how the time evolved Monte Carlo wave-functions can be used to yield stochatic "process vectors" $\zeta$  that directly average to the process matrix $\chi$. In Sec. VI, we show how the no-jump, single wave function trajectory yields a readily accessible upper bound on the trace distance between the actual process matrix and any desired process matrix - a useful measure of process fidelity. In Sec. VII, we apply our method to the simulation of a Rydberg blockade C-PHASE gate, subject to realistic decay and dephasing mechanisms. Sec. VIII concludes the paper.     

\section{\label{sec:oqs} Characterising Open Quantum Systems}
The simulation of dissipative quantum systems is important to quantum information processing as it allows realistic modelling of quantum gate protocols. Because of dissipation, the time-evolution is not unitary and the dynamics must be treated using a master-equation approach. This linear equation, describing time evolution of the principal system's density matrix $\rho$, is typically obtained by making use of the Born-Markov approximations and tracing a larger composite density matrix over the reservoir degrees of freedom associated with the dissipation processes. Denoting the Hamiltonian for the principal system by $H$, the master equation may be written as
\begin{equation}
\dot{\rho} = -\frac{i}{\hbar}[H,\rho]+\mathcal{L}_{\rm relax}(\rho)\, , \label{eq:ME}
\end{equation}
where we consider the relaxation operator $\mathcal{L}_{\rm relax}$ in Lindblad form~\cite{gardiner2004quantum}
\begin{equation}
\mathcal{L}_{\rm relax}(\rho)=-\frac{1}{2}\sum_k (L_k^{\dag}L_k\rho + \rho L_k^{\dag}L_k) + \sum_k L_k\rho L_k^{\dag}\, .
\label{eq:lindblad}
\end{equation}
Eq.(1) preserves the positivity and normalization of the density operator $\rho$ and the $L_k$ operators in Eq.~(\ref{eq:lindblad}) act in the space of the principal system. Thus by solving Eq.~(\ref{eq:ME}) with the, possibly time-dependent, Hamiltonian corresponding to a complex quantum gate operation we determine the evolution of any initial state under the influence of damping and noise.

More generally, the net effect of a quantum operation in an open quantum system may be described in the operator-sum representation~\cite{kraus1983states}
\begin{equation}
\mathcal{E}(\rho)=\sum_i K_i\rho K_i^{\dag}\, ,
\label{eq:opsum}
\end{equation}
where the Kraus operators $K_i$ act on the system's Hilbert space and obey $\sum_i K_i^{\dag}K_i = 1$. Picking the Hermitian operators $\{E_m\}$ as a basis for the set of all operators on the principal system's Hilbert space~\footnote{In a multi-qubit system, an appropriate operator basis might be tensor products of the identity and Pauli operators for each qubit.} we may write any quantum process as
\begin{equation}
\mathcal{E}(\rho)=\sum_{mn}\chi_{mn} E_m\rho E_n^{\dag}\, .
\label{eq:fixbasis}
\end{equation}
The characterisation matrix $\chi_{mn}$ in Eq.~(\ref{eq:fixbasis}) is related to the Kraus form~(\ref{eq:opsum}) via the expansion of each $K_i=\sum_m e_{im}E_m$, and the identification $\chi_{mn}=\sum_{i}e_{im}e_{in}^{\ast}$. 

Suppose a process is simulated with the master equation, for which the accumulated effect of the unitary and dissipative dynamics on the quantum system is not known a priori. From the simulation data, we want an efficient method to obtain the full information about $\mathcal{E}$. In the next section standard methods of acquiring $\chi_{mn}$, and thus $\mathcal{E}$, will be shown.

\section{\label{sec:estchar} Process Characterisation Schemes}
In the following discussion we consider a general quantum system with Hilbert space dimension $D$. Assuming the map $\mathcal{E}$ is trace preserving then characterisation of $\mathcal{E}$ is equivalent to a determination of the $D^4-D^2$ independent elements of $\chi$~\cite{nielsen2010quantum}.

Let $\mathcal{O}_{pq}=|p\rangle\langle q|$ for $p,q\in\{1,\dots,D\}$ be a linearly independent basis for the space of $D\times D$ linear operators. Cataloguing the action of the fixed operator basis $\{E_m\}$ on all input matrices we create the $D^4\times D^4$ matrix $\mathcal{K}$:
\begin{equation}
E_m\mathcal{O}_{rs}E_n^{\dag}=\sum_{pq}\mathcal{K}_{rs,pq}^{mn}\mathcal{O}_{pq}\, .
\label{eq:beta}
\end{equation}

\subsection{Standard quantum process characterisation}
In standard quantum process characterisation (SQPC) the effect of the process $\mathcal{E}$ is determined: Experimentally, $D^2$ different input states (density matrices) are subjected to the physical process and the resulting output states are measured by quantum state tomography. In a theoretical analysis, the master equation is used to simulate the process and the outcome solution $\mathcal{E}(\mathcal{O}_{rs})$ for input matrices $\mathcal{O}_{pq}$ is expressed as a linear combination in the same operator basis,
\begin{equation}
\mathcal{E}(\mathcal{O}_{rs})=\sum_{pq}\Lambda_{rs,pq}\mathcal{O}_{pq}\, .
\label{eq:lambda}
\end{equation}

Combining Eqs.(4-6) we obtain $\sum_{mn}\mathcal{K}_{rs,pq}^{mn}\chi_{mn}=\Lambda_{rs,pq}$, which in matrix form reads
\begin{equation}
\boldsymbol{\mathcal{K}\chi}=\boldsymbol{\Lambda}\, .
\label{eq:lineqn}
\end{equation}
Finding $\{\chi_{mn}\}$ from the simulated $\{\Lambda_{rs,pq}\}$ is now a linear algebra problem, although in general it is not uniquely determined by Eq.~(\ref{eq:lineqn}).

Let us here make an estimate of how the computational resources needed to perform SQPC scales with Hilbert space dimension. Simulating a process with the master equation requires solving $D^2$ coupled differential equations for each of the $D^2$ input states; that is, we must solve $D^4$ differential equations. Solution of Eq.~(\ref{eq:lineqn}) requires decomposition of $\mathcal{K}$, using the Cholesky method for example, followed by forward and back substitution for $\chi$. The computational complexity of a straightforward decomposition is O($D^{12}$) while the substitution operations are each O($D^8$). 

For applications to quantum computing on a register composed of $N$ L-level quantum systems, the product Hilbert space has the dimension $D=L^N$. If the operators $E_m$ are taken to be SU(L) operator products, the product nature simplifies decomposition of $\mathcal{K}$ into separate O($L^{12}$) problems.

\subsection{Ancilla assisted process characterisation}
In ancilla assisted process characterisation (AAPC), instead of composing $\Lambda$ by propagating separate, initial matrices $\mathcal{O}_{rs}$, all input states are simultaneously represented in a ``super'' operator
\begin{equation}
\label{eq:superop}
\mathcal{O}=\sum_{rs} \mathcal{O}_{rs}\otimes \mathcal{O}_{rs}\, .
\end{equation}
on the combined principal ($P$) and ancilla ($A$) system. This expanded system is now made subject to the quantum process, $\mathcal{E}_{P\otimes A}(\mathcal{O})\to\mathcal{O}_{\rm out}$, which acts with the original process $\mathcal{E}$ only on the principal system component, 
\begin{equation}
\label{eq:choi}
\mathcal{O}_{\rm out}\equiv(\mathcal{E}\otimes \mathcal{I})(\mathcal{O})\, .
\end{equation}
The identity operator $\mathcal{I}$ in Eq.~(\ref{eq:choi}) acts on the ancilla's operator space. From Eq.~(\ref{eq:superop}) we have $(\mathcal{E}\otimes\mathcal{I})(\mathcal{O})= \sum_{rs} \mathcal{E}(\mathcal{O}_{rs})\otimes \mathcal{O}_{rs}$, implying that a single master equation simulation on the expanded system allows calculation of all $\mathcal{E}(\mathcal{O}_{rs})$. The ancilla system is used to extract separate results,
\begin{equation}
\mathcal{E}(\mathcal{O}_{rs})={\rm Tr}_A\big[(I\otimes |s\rangle\langle r|)\mathcal{E}_{P\otimes A}(\mathcal{O})\big]\, ,
\end{equation}
where ${\rm Tr}_A$ denotes partial trace on the ancilla's Hilbert space. Then in a way equivalent to Eq.~(\ref{eq:lambda}) we may expand $\mathcal{E}(\mathcal{O}_{rs})$ into the basis of $\{\mathcal{O}_{pq}\}$ and retrieve the characterisation matrix $\chi$ from Eq.~(\ref{eq:lineqn}).

Simulating a quantum process with AAPC involves solving $D^4$ differential equations for the expanded input state. The complexity of solving Eq.~(\ref{eq:lineqn}) remains O($D^{12}$) and O($D^{8}$) for decomposition of $\mathcal{K}$ and forward/backward substitution respectively.

\section{Monte-Carlo Wave-Functions}
We may obtain the predictions made by the master equation~(\ref{eq:ME}) by introducing a stochastic element into the evolution of so-called Monte-Carlo wave-functions~\cite{PhysRevLett.68.580,Molmer:93,PhysRevA.45.4879}. These are wave-functions $|\psi(t)\rangle$ propagated with the non-Hermitian Hamiltonian $H_{\rm eff}=H-i\hbar/2\sum_{k}L_k^{\dag}L_k$. Due to the non-unitary evolution during a small time step $dt$
\begin{equation}
|\psi_0(t+dt)\rangle = \left (1+\frac{1}{i\hbar}H_{\textrm{eff}} dt \right )|\psi(t)\rangle\, ,
\label{eqn:waveevo}
\end{equation}
the square of the norm associated with $|\psi_0(t+dt)\rangle$ is reduced by
\begin{equation}
\delta p = \sum_k \delta p_k = dt\sum_k \langle\psi(t)|L_k^{\dag}L_k|\psi(t)\rangle\, .
\label{exp:littlep}
\end{equation}
The next step involves a random choice. Either the wave function $|\psi_0(t+dt)\rangle$ is re-normalised or, with probability $\delta p$, the wave function is subject to a quantum jump. This constitutes a collapse of the wave-function and with a branching ratio of $\delta p_k/\delta p$, the final state is chosen among the states $L_k|\psi(t)\rangle$. Thus, at time $t+dt$ we have one of the following possibilities:
\begin{subequations}
\begin{align}
\text{with prob. }& 1-\delta p,\quad |\psi(t+dt)\rangle = \frac{|\psi_0(t+dt)\rangle}{\sqrt{1-\delta p}}\, ;\label{eqn:probevo1} \\
\text{with prob. }& \delta p_k,\quad |\psi(t+dt)\rangle = \frac{L_k|\psi(t)\rangle}{\sqrt{\delta p_k/dt}}\, .
\label{eqn:probevo2}
\end{align}
\end{subequations}
An ensemble of wave fucnctions subject to this dynamics will on average reproduce the time dependent solution of the master equation. 

The validity of the calculation relies on time steps much smaller than the time-scale of the coherent and incoherent physical processes. However, the direct implementation discussed above may be reformulated ~\cite{PhysRevA.45.4879,PhysRevLett.72.203} such that  quantum jumps are not decided in terms of expression~(\ref{exp:littlep}), linear in the ``small'' time step $dt$. Instead, the ``no-jump'' wave-function $|\psi_0(t)\rangle$ is allowed to evolve until its norm reaches a predetermined random number uniformly distributed between 0 and 1. At this time a jump is made in the manner of Eq.~(\ref{eqn:probevo2}). In this way  integration of Eq.~(\ref{eqn:waveevo}) may be left to an accurate and efficient numerical solver. In our implementation to a physical example below, we use a variant of the Adams-Bashford method which utilizes adaptive step-size control.

\section{\label{sec:montchar} Process characterisation with Monte-Carlo}
Since density matrices in the master equation approach are replaced by state vectors, simulation by Monte-Carlo wave-functions allows access to larger quantum systems. How this benefits process characterisation is the focus of this section. We first analyse ancilla assisted process characterisation using Monte-Carlo wave-functions (AAWF) and discuss its advantages over standard characterisation techniques. We conclude the section with a discussion of why the ancilla strategy is preferred over a Monte-Carlo treatment of standard quantum process tomography.

\begin{figure}
\includegraphics[width=0.95\columnwidth]{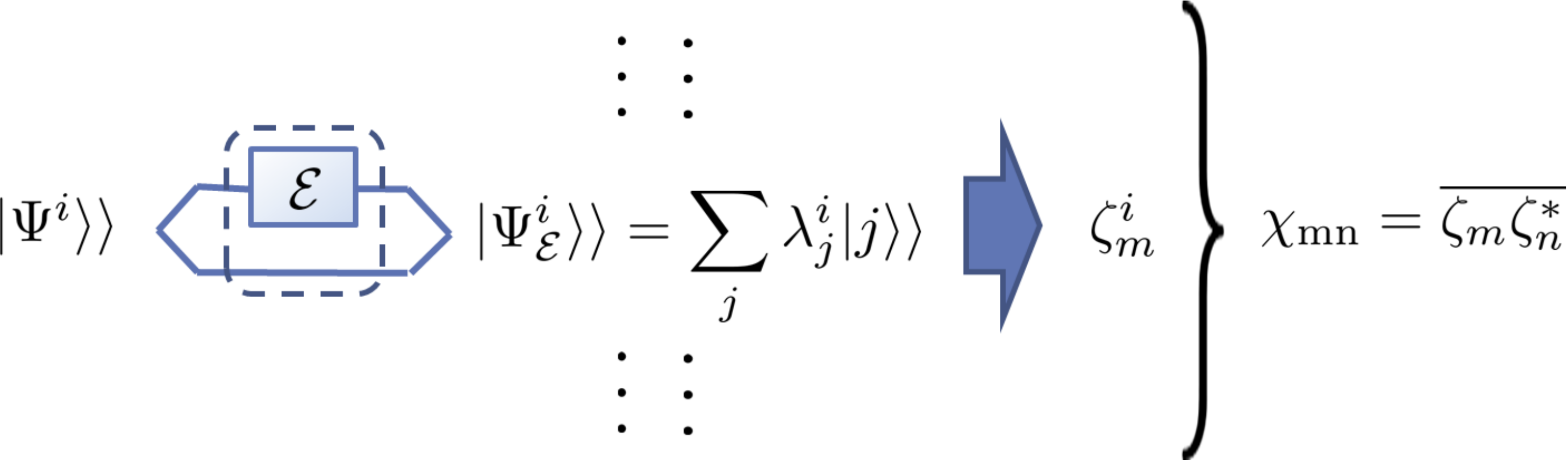}
\caption{\label{fig:DWPC}Implementation of ancilla assisted process characterisation with Monte-Carlo wave-functions: In each trajectory $i$, the evolution of the entangled system and ancilla is simulated, subject to the process $\mathcal{E}\otimes I$.  From the expansion coefficients $\lambda_j$, we determine the vector of components $\zeta_m$. Averaging over many simulation outcomes $\zeta_m^{i}\zeta_n^{i\ast}$ the $\chi$-matrix is obtained. }
\end{figure}

Observe that the super operator in Eq.~(\ref{eq:superop}) is a pure state projection operator, that is $\mathcal{O}=|\Psi\rangle\rangle\langle\langle\Psi|$, where the bipartite wave-function $|\Psi\rangle\rangle=\sum_{r} |r\rangle\otimes |r\rangle$ describes the principal system maximally entangled with an identical ancilla system. Evolution of $\mathcal{O}$ is then equivalent to propagation of $|\Psi\rangle\rangle\to|\Psi_{\mathcal{E}}\rangle\rangle$ under the rules of Monte-Carlo wave-functions. That is, averaging the outer product of the resulting states over many simulated outcomes we find $\overline{|\Psi_{\mathcal{E}}\rangle\rangle\langle\langle\Psi_{\mathcal{E}}|}=(\mathcal{E}\otimes \mathcal{I})(\mathcal{O})$.

Next we construct $\kappa$, the ```wave-function'' analogue of standard tomography's $\mathcal{K}$-matrix. This is a $D^2\times D^2$ matrix matrix of expansion coefficients for the application of the standard operators $\{E_m\}$ of section~\ref{sec:oqs},
\begin{equation}
(E_m\otimes I)|\Psi\rangle\rangle=\sum_j \kappa_j^m |j\rangle\rangle\, ,
\label{eq:kappa}
\end{equation}
where $\{|j\rangle\rangle:j=1,\dots,D^2\}$ is a linearly independent basis for the $D^2$-dimensional vectors on the principal-ancilla space $P\otimes A$. 

For each simulated Monte-Carlo wave-function $|\Psi_{\mathcal{E}}\rangle\rangle=\sum_{j}\lambda_{j}|j\rangle\rangle$ we now define the $\zeta$-vector as the solution to
\begin{equation}
\label{eq:linearzeta}
\sum_m\kappa_j^m\zeta_m=\lambda_j\, .
\end{equation}
The coefficients $\lambda_j$, and hence $\zeta_m$, carry information about the process $\mathcal{E}$. Indeed, when averaging the $\zeta$-vector coordinates over a sufficiently large ensemble of Monte-Carlo wave-functions we directly obtain the process matrix of Eq.~(\ref{eq:fixbasis}),
\begin{equation}
\chi_{\rm mn}=\overline{\zeta_m\zeta_n^{\ast}} ,
\end{equation}
(see appendix).

The implementation of the AAWF method is illustrated in Fig.~\ref{fig:DWPC}. To propagate $|\Psi\rangle\rangle$ in a single Monte-Carlo trajectory we must solve $D\times D$ coupled differential equations. Thus, averaging over $n$ trajectories requires solving $nD\times D$ coupled differential equations, which may be much less than the $D^4$ equations needed by standard techniques. Even more striking is the reduction in cost associated with the decomposition of $\kappa$, which scales as a O($D^6$) problem, compared to the O($D^{12}$) problem of decomposing the $\mathcal{K}$-matrix. Each of the $n$ Monte-Carlo simulations at output creates a size $O(D^4)$ problem when using forward/backward substitution to solve for the $\zeta$-vector.

Analogous to standard process characterisation, we might have used Monte-Carlo wave-functions to simulate input states in the original Hilbert space, construct the output density matrices and subsequently solve for $\chi$. However, implementation requires simulating $n$ size $D$ Monte-Carlo wave-functions for all $D^2$ input states, which is a larger problem than its AAPC counterpart (Table~\ref{tab:PTstrategies}). Also, this method does not allow for a $\zeta$-vector equivalent, meaning no reduction in the complexity of solving Eq.~(\ref{eq:lineqn}). 

Note that the decomposition of $\mathcal{K}$ concerns only structural properties of the chosen operator and state bases of the quantum system and is independent of the physical process $\mathcal{E}$. The same is true for the vector variant, applied to the analysis of Monte-Carlo wave functions. Indeed, our vector formulation of the problem shows that the simpler decomposition of $\kappa$ offers an effective reduction of the costs to decompose $\mathcal{K}$, which is applicable for standard process characterisation.

\begin{table}
\caption{\label{tab:PTstrategies}Numerical cost of characterising a quantum process on a $D$ dimensional Hilbert space. The first column lists the density matrix and Monte-Carlo approaches to standard (SQPC) and ancilla assisted process characterisation (AAPC). The second column (c.d.e.) lists the number of coupled differential equations needed to simulate the time evolution. The last column (s.l.e.) lists the cost of solving the system of linear equations for the $\chi$-matrix/$\zeta$-vector, assuming a Cholesky decomposition of the matrix  $\mathcal{K}$ ($\kappa$).}
\begin{ruledtabular}
\begin{tabular}{lcc}
Technique & \ c.d.e. &  s.l.e. \\ \hline
SQPC: &  &  \\
\quad\quad density matrix & $D^2\times D^2$  & $O(D^8)$ \\
\quad\quad Monte-Carlo & $nD^2\times D$  & $O(D^8)$ \\
AAPC: &  &  \\
\quad\quad density matrix & $D^2\times D^2$  & $O(D^8)$ \\
\quad\quad Monte-Carlo & $nD\times D$  & $n\times O(D^4)$ \\
\end{tabular}
\end{ruledtabular}
\end{table}

\section{\label{sec:uppbnd}An upper bound error estimate from a single wave function}
A meaningful measure between the ideal ($\tilde{\chi}$) and the actual ($\chi$) process matrices in a physical gate operation is the trace distance ~\cite{PhysRevA.71.062310}. The trace distance is a translationally invariant metric on the space of Hermitian, positive semi-definite matrices of unit trace, and it is given by $T(\tilde{\chi},\chi) \equiv\frac{1}{2}\lVert\tilde{\chi}-\chi\rVert_{tr}$, where $\lVert A \rVert_{tr}={\rm Tr}(\sqrt{A^{\dag}A})$ is the trace norm.

We are interested in the characterisation of the effect of noise and dissipation on gate operations in quantum computing. To have any relevance for quantum computation such gates may only experience weak noise. This implies that in the majority of Monte-Carlo simulations, the wave functions should follow the ``no-jump'' trajectory~(\ref{eqn:probevo1}) through the entire duration of the process. With the AAWF method, the calculation of this single wave function provides a useful indication of the gate's performance.

The simulated Monte-Carlo wave-functions making up a AAWF calculation of $\chi$ may be separated into two parts; those that never jumped, yielding a single $\zeta$-vector and corresponding $\chi$-matrix, $\chi_S=\zeta_S\zeta_S^{\dag}$ and those that jumped at randomly assigned times to yield a set of vectors ($\chi_i=\zeta_i\zeta_i^{\dag}$). The fraction $\frac{S}{n}$ of no-jump wave-functions is equal to the product of the normalisation factors ($1-\delta p$) in Eq.~(\ref{eqn:probevo1}) applied over time. The remaining fraction $\frac{J}{n}$ of the simulated ensemble ($n=J+S$ being the total number of trajectories) yields the sum of terms $\chi_J= \frac{1}{J}\sum_{i=1}^J \zeta_i\zeta_i^{\dag}$. Finally we see that
\begin{equation}
\label{eq:separate}
\chi = \frac{S}{n}\chi_S + \frac{J}{n}\chi_J\, .
\end{equation}

Calculating the trace distance between Eq.~(\ref{eq:separate}) and the ideal process matrix $\tilde{\chi}$ we employ the triangle inequality and translational invariance to obtain
\begin{equation}
T(\tilde{\chi},\frac{S}{n}\chi_S+\frac{J}{n}\chi_J)\le T(\tilde{\chi}-\frac{S}{n}\chi_S,0)+T(0,\frac{J}{n}\chi_J).
\end{equation}
Since $T(0,\frac{J}{n}\chi_J)=\frac{1}{2}\lVert\frac{J}{n}\chi_J\rVert_{tr}=\frac{J}{2n}$, this provides an upper bound on the trace distance using the evolution of only a single no-jump wave-functions and its associated $\chi$ matrix,
\begin{equation}
T(\tilde{\chi},\chi_S+\chi_J)\le T(\tilde{\chi},\frac{S}{n}\chi_S)+\frac{J}{2n}.
\end{equation} 
Clearly, this upper bound is of limited value if dissipation is significant and many wave functions jump.

Another typical measure for the effect of error in a process is fidelity $F(\tilde{\chi},\chi)\equiv \lVert\sqrt{\tilde{\chi}}\sqrt{\chi}\rVert_{tr}$. However, being a non-linear expression it has a more complicated relation with the different components of the wave-function ensemble. We shall return to both process error measures in the numerical example below.

\begin{figure}
\includegraphics[width=0.95\columnwidth]{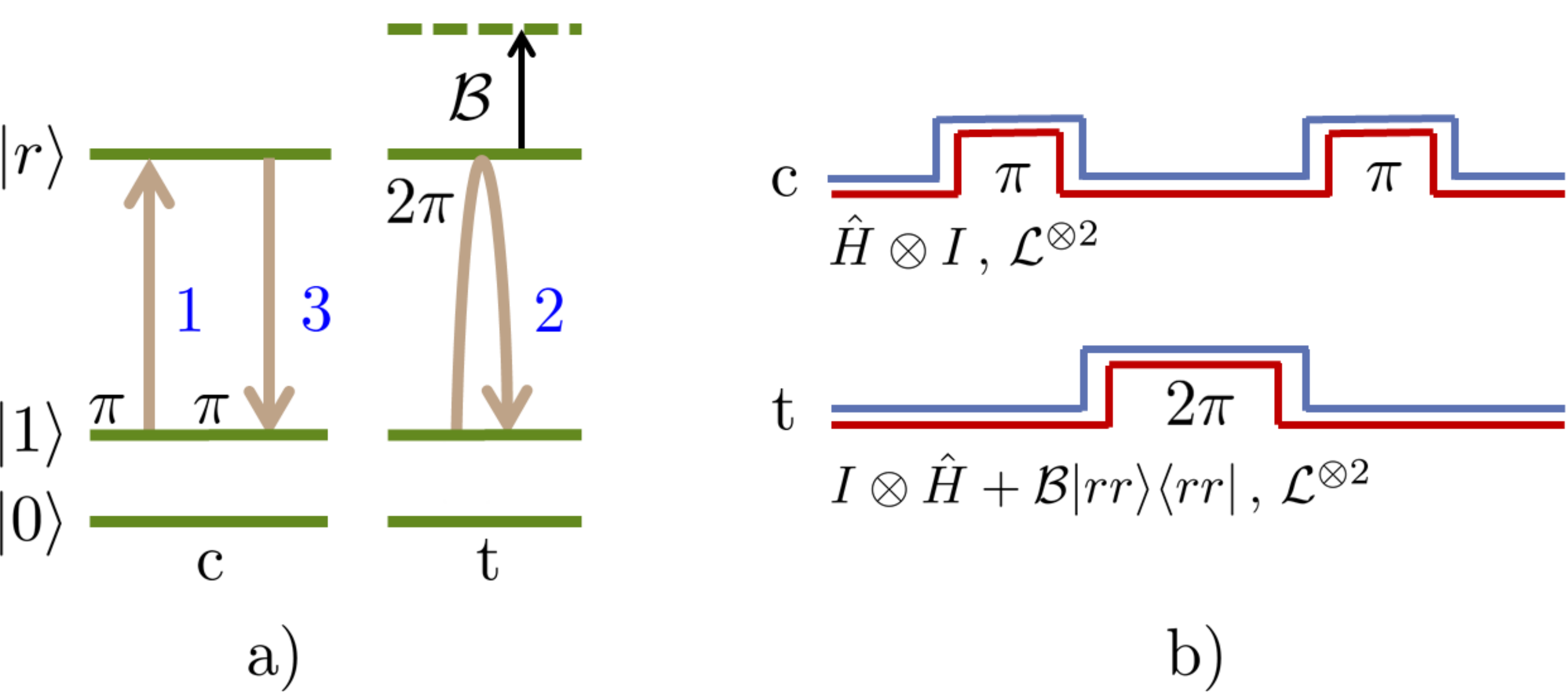}
\caption{\label{fig:cphase}Rydberg mediated controlled phase gate. Part a) of the figure illustrates 1: a resonant transfer between $|1\rangle$ and $|r\rangle$ of the control atom $c$. 2: subsequent coherent excitation and de-excitation between $|1\rangle$ and $|r\rangle$ of the target atom $t$, yielding the controlled phase shift on the $|1\rangle$ component, unless the control atom is excited and provides the Blockade shift $\cal{B}$. 3: de-excitation of the control atom.  Part b) illustrates the  pulse sequence and lists the Hamiltonian and driving terms. The identity operator $I$ signifies individual addressing of atoms, $\mathcal{L}^{\otimes 2}=-\frac{i\hbar}{2}(\sum_{m}L_m^{\dag}L_m\otimes I + I\otimes\sum_{m}L_m^{\dag}L_m)$ describes decoherence in the system and the two-atom interaction $\mathcal{B}|rr\rangle\langle rr|$ prevents both atoms from occupying the Rydberg state. The single qubit Hamiltonian $\hat{H}$ is discussed in detail in the text.}
\end{figure}

\section{C-PHASE Gate}
As an example of our process characterisation, we consider the controlled-phase (C-PHASE) gate operation between atoms coupled by the Rydberg blockade interaction. Each atom has four levels; two ground levels which comprise the qubit space, the Rydberg level with which the atoms interact, and the intermediate level which facilitates transitions to the Rydberg level via a two-photon process. Thus, even though the initial state space and resulting process matrix may be restricted to the qubit space we are still required to simulate the system considering all levels. Obtaining the $\chi$-matrix even for the simplest two-qubit gates is non-trivial, and extending it beyond two atoms becomes computationally challenging for standard characterisation strategies. Meanwhile, the proposed AAWF method can readily deal with multi-qubit process characterisation involving up to 8 or 10 atoms.

A C-PHASE gate on the 5s$_{1/2}$ hyperfine states $|0\rangle\equiv|F=1,m_F=0\rangle$ and $|1\rangle\equiv|F=2,m_F=0\rangle$ involves single qubit rotations between $|1\rangle$ and the Rydberg state $|r\rangle=|97d_{5/2},m_j=5/2\rangle$. This is a two photon process, achieved with $\sigma_+$ polarised 780- and 480-nm beams. The 780-nm beam is tuned by an amount $\Delta$ to the red of the $|1\rangle\to|p\rangle\equiv|5p_{3/2},F=3\rangle$ transition while the 480-nm beam is also tuned an amount $\Delta$ to the blue of the $|p\rangle\to|r\rangle$ transition. The resulting Rabi frequencies are $\Omega_R$ ($\Omega_B$) for the red (blue) detuned laser. After adiabatically eliminating $|p\rangle$ from the Hamiltonian describing this process for numerical efficiency, we find in the rotating-wave approximation~\cite{PhysRevA.85.032111}
\begin{eqnarray}
\label{eq:effham2}
\hat{H} =&& \underbrace{\frac{(2\Delta-\delta E_r)\Omega_R\Omega_B}{8\Delta(\Delta-\delta E_r)+2\gamma^2}}_{\Omega_{\rm eff}/2}|1\rangle\langle r| + {\rm H.c.} \nonumber\\
&&\, + \left(\delta E_0-\frac{\Delta\Omega_R^2}{4\Delta^2+\gamma^2}\right)|0\rangle\langle 0|\, .
\end{eqnarray}
Here
\begin{equation}
\delta E_r\simeq\frac{16\Delta^2(\Omega_B^2-\Omega_R^2)-\Omega_R^4}{64\Delta^3}\,
\end{equation}
is subtracted from the ``blue'' detuning to compensate for sub-optimal Rabi-oscillations due to Stark shifts arising from power differences between the red and blue detuned lasers. Similarly, $\delta E_0$ may be used to balance the $|0\rangle\langle 0|$ term, ensuring no phase contributions from $|0\rangle$ caused by the red laser. Finally, $\gamma$ is the decay rate from $|p\rangle$.

\begin{figure}
\includegraphics[width=0.75\columnwidth]{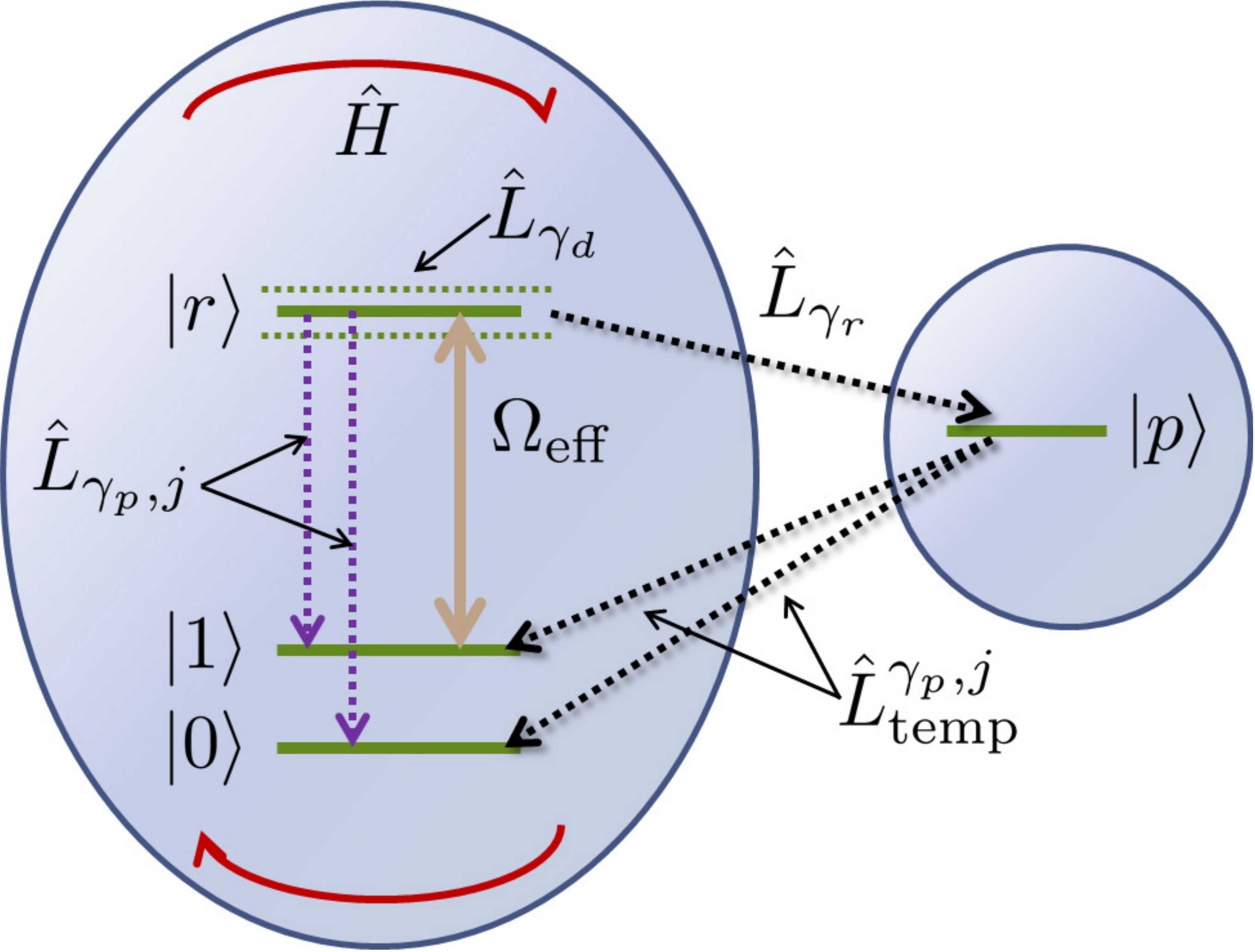}
\caption{\label{fig:effsys} For numerical efficiency the shirt-lived, intermediate state $|p\rangle$ is adiabatically eliminated and yields an effective description with couplings ($\Omega_{\rm eff}$) and dissipation terms ($\hat{L}_{\gamma_d}$,$\hat{L}_{\gamma_p ,j}$) shown in the left part of the figure. For convenience, $|p\rangle$ is formally reintroduced to also represent decay events from $|r\rangle$ (see text).}
\end{figure}

The effective operator formalism~\cite{PhysRevA.85.032111} provides us with a mechanism to simulate decay from $|p\rangle$, viz.
\begin{equation}
\hat{L}_{\gamma_p ,j} = \frac{\sqrt{c_j\gamma_p}\,\Omega_R}{2\Delta-i\gamma}|j\rangle\langle 1| + \frac{\sqrt{c_j\gamma_p}\,\Omega_B}{2(\Delta-\delta E_r)-i\gamma}|j\rangle\langle r|\, ,
\end{equation}
where $j\in\{0,1,g\}$ and $c_j$ are the branching ratios $\{0.12,0.32,0.56\}$ for decay from $|p\rangle$, that is, $\gamma=\sum_jc_j\gamma_p$. It is appropriate here to discuss $\hat{L}_{\gamma_p ,g}$ because $|g\rangle$ does not feature in the system Hamiltonian. Decay events into the ``loss state'' $|g\rangle$ do not couple back into the system, and a Monte-Carlo trajectory is merely disposed when a jump of this sort is simulated. 

Magnetic field noise and atomic motion are important dephasing sources that we describe in Monte Carlo simulation by the jump operator~\cite{Molmer:93}
\begin{equation}
\hat{L}_{\gamma_d} = \sqrt{\gamma_d}(\mathbb{1}-2|r\rangle\langle r|)\, ,
\end{equation}
where $\mathbb{1}$ is shorthand for the identity operator. Spontaneous emission from $|r\rangle$, described by the jump operator
\begin{equation}
\hat{L}_{\gamma_r} = \sqrt{\gamma_r}|p\rangle\langle r|\, ,
\end{equation}
populates the eliminated state $|p\rangle$. Temporarily reintroducing $|p\rangle$ at jump times, followed by immediate jumps to the lower lying states by $\hat{L}_{\rm temp}^{\gamma_p ,j} = \sqrt{c_j\gamma_p}|j\rangle\langle p|$, allows accurate simulation of the decay processes. The parameters chosen for our simulations are summarized in Table II.
 
\begin{table}
\caption{\label{tab:params}Physical parameters for our simulations based on values discussed in Refs.~\cite{PhysRevLett.100.113003,saffman2011rydberg}.}
\begin{ruledtabular}
\begin{tabular}{lcr}
Experimental parameter & Symbol & Value \\ \hline
Detuning & $\Delta/2\pi$ & 2.0 GHz \\
Red Rabi frequency & $\Omega_R/2\pi$ & 118 MHz \\
Blue Rabi frequency & $\Omega_B/2\pi$ & 39 MHz \\
Rydberg blockade & $\mathcal{B}/2\pi$ & 20 MHz \\
Decay rate ($|p\rangle$) & $\gamma_p/2\pi$ & 6.07 MHz \\
Decay rate ($|r\rangle$) & $\gamma_r/2\pi$ & 0.53 kHz \\
Dephasing rate ($|r\rangle$) & $\gamma_d/2\pi$ & 1.0/2.0 kHz \\
\end{tabular}
\end{ruledtabular}
\end{table}

\begin{figure}
\includegraphics[width=\columnwidth]{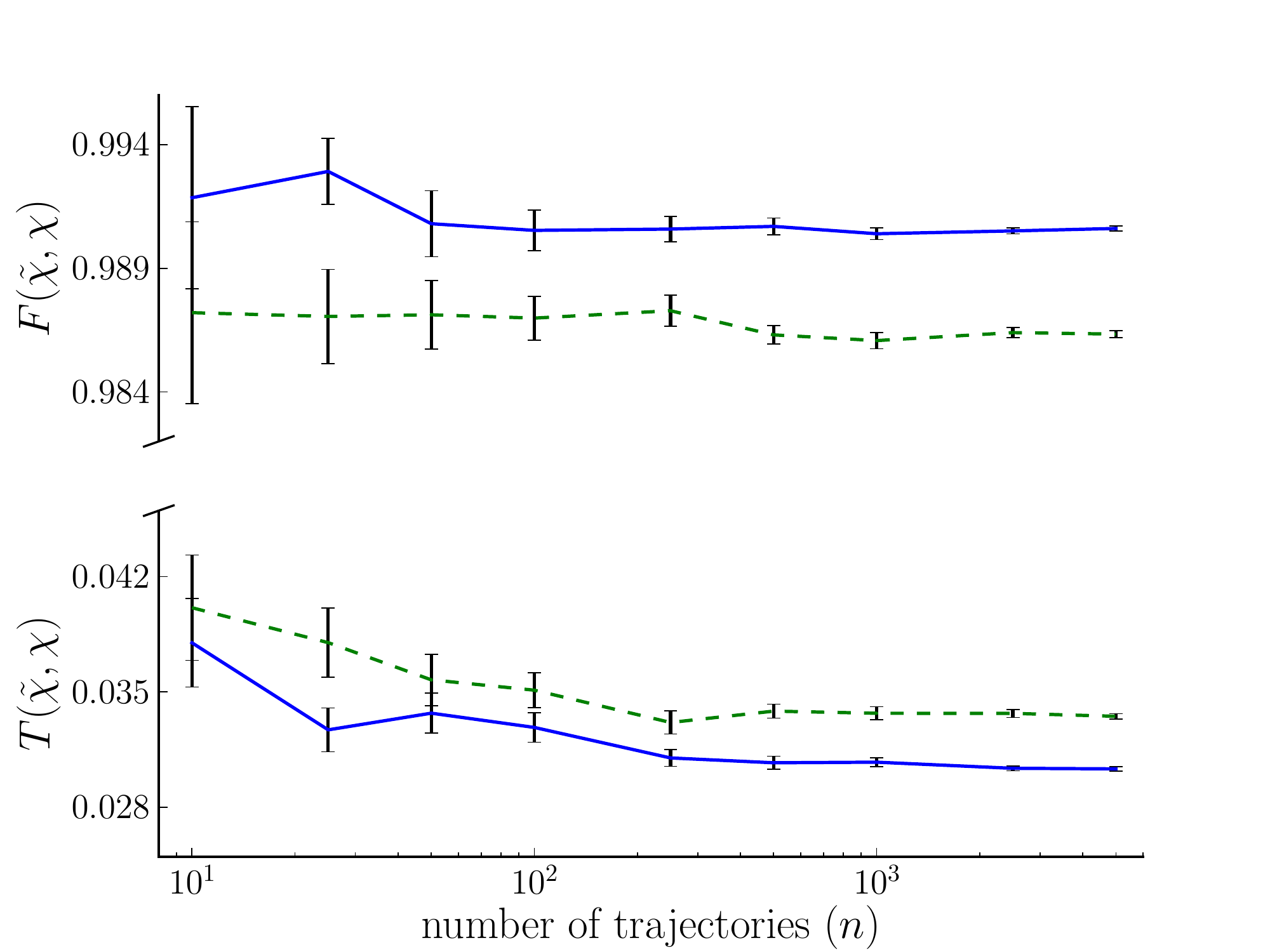}
\caption{\label{fig:converge}Convergence of the process fidelity and trace distance determined by the ancilla assisted wave function characterisation method. Mean values and standard deviations of the fidelity (top) and trace-distance (bottom) are obtained with 50 samples at each value of $n$. The solid  lines represent calculations using the parameters in Table II with $\gamma_d/2\pi=$ 1.0 kHz. The dashed lines are obtained with the higher dephasing rate $\gamma_d/2\pi=$ 2.0 kHz.}
\end{figure}

The Monte Carlo wave functions on average yield the density matrix. This is true for any ensemble size, while the statistical errors on the estimate decrease with large $n$. The trace distance and the fidelity measures are not linear functions in the density matrix elements. Hence sampling their values with a finite wave function ensemble may provide a systematic error in addition to the statistic uncertainty of the method. In~\cite{PhysRevA.54.5275}, a non-linear master equation was analysed and the systematic error was estimated to scale as $1/n$, thus becoming less important than the statistical error ($\sim 1/\sqrt{n}$) for large ensembles. Convergence of the AAWF method is illustrated in Fig.~\ref{fig:converge} where trace distance $T(\tilde{\chi},\chi)$ and fidelity $F(\tilde{\chi},\chi)$ are recorded for different Monte-Carlo wave-function ensemble sizes. For each ensemble size, we have made 50 simulations and at $n=500$ sample-to-sample variations are small enough to consider the output results satisfactorily converged.

\begin{figure}
\includegraphics[width=\columnwidth]{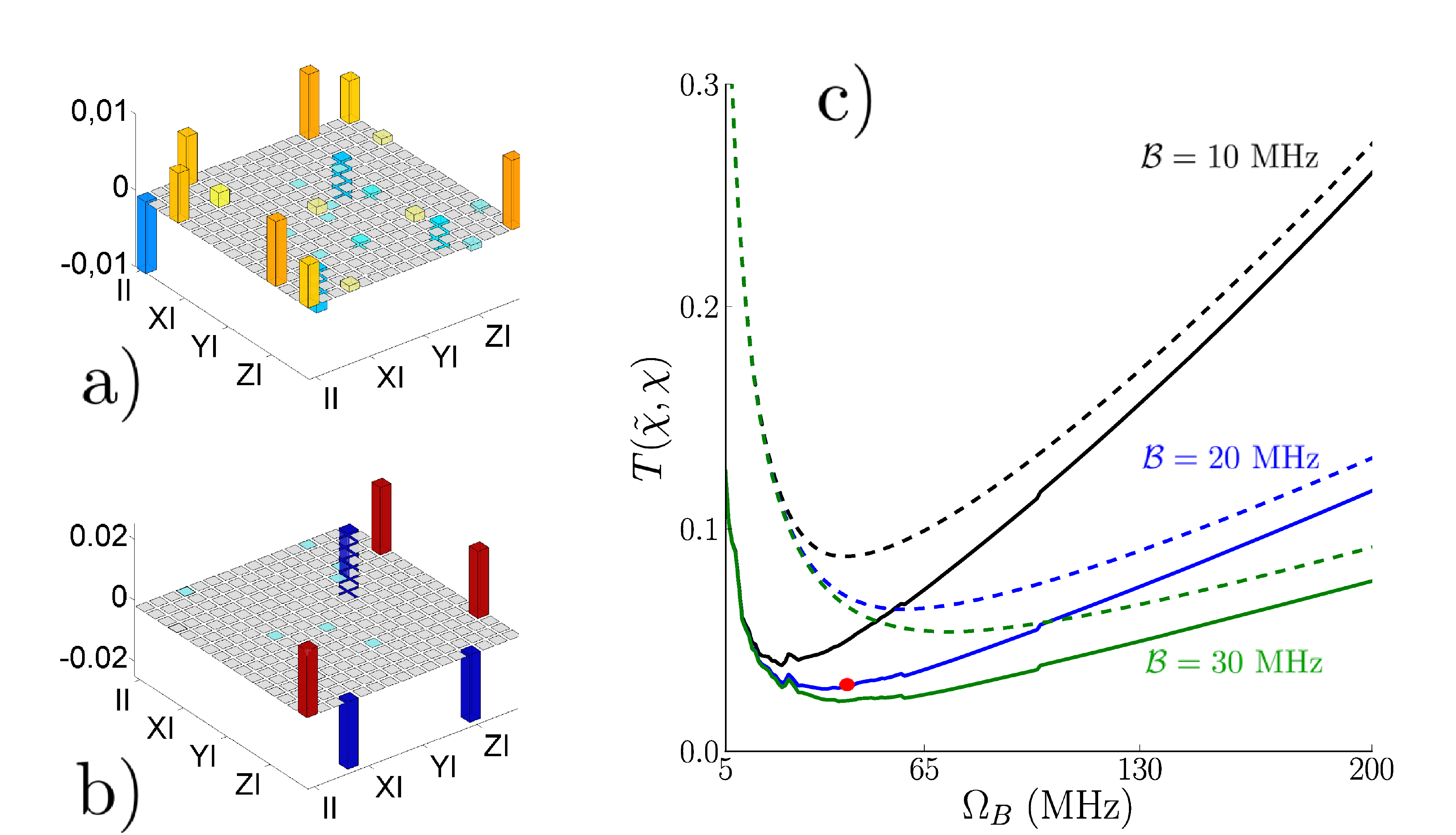}
\caption{\label{fig:upabnd} Real and imaginary part of the difference between the actual and the desired process matrix, a) and b), for a C-PHASE gate [N.B. Sub-tick labels to the right of each $\mathcal{W}$I read $\mathcal{W}$X, $\mathcal{W}$Y, $\mathcal{W}$Z, where $\mathcal{W}\in \{I,X,Y,Z\}$.] Part c) of the figure shows the trace distance as a function of the blue laser Rabi frequency for three experimental realisations of blockade strength: full AAWF treatment (solid line) and no-jump wave function upper bound (dotted line). The (red) dot on the solid $\mathcal{B}=20$ MHz line indicates the parameters used in parts a) and b) of the figure. We use the parameters listed in Table II with $\gamma_d/2\pi=$ 1.0 kHz.}
\end{figure}

In Figs.~\ref{fig:upabnd}.a and b, we show the real and imaginary part of the difference between the process matrix elements obtained by our simulations and the ideal C-PHASE gate. Fig.~\ref{fig:upabnd}.c shows the trace distance between the gates as function of the blue laser Rabi frequency for different values of the Rydberg blockade shift. The solid curves are based on our simulations with ensembles of $n=500$ Monte-Carlo wave-functions, while the dashed curves are upper bound calculations using a single no-jump trajectory for each set of parameters, cf. Sec.~\ref{sec:uppbnd}. The figure confirms that the upper bound indeed exceeds simulation results and, given its simplicity, provides a reasonable characterisation of the errors. To understand the variation of the trace distance for small values of $\Omega_B$ we recall that the gate time $t_{\rm gate}\propto 1/\Omega_{\rm eff}$. As $\Omega_B\to 0$ the gate time lengthens and errors due to intermediate state decay and dephasing increase. On the other hand, as $\Omega_{\rm eff}\simeq \Omega_B\Omega_R/\Delta\to\mathcal{B}$ from below, the gate errors increase due to population leakage into the $|rr\rangle$ state. We thus find an optimum value for $\Omega_B$ between these two regimes. Although certain to pose experimental challenges, the simulation also records advantages to gate quality by increasing $\mathcal{B}$ to 30 MHz. 

\section{Discussion and Conclusion}
In conclusion we have presented a numerically efficient method to obtain the $\chi$-matrix for an arbitrary quantum process from a solution to the system's master equation. Monte-Carlo wave-functions present an effective means to simulate the system density matrix and extends in a natural way to model ancilla-assisted process characterisation. Parametrizing the outcomes of the simulated system under a fixed set of operations we presented a $\zeta$-vector representation of the Monte-Carlo wave-functions. Because the process matrix $\chi$ results as a simple product of $\zeta$-vector components the numerical effort to both simulate state evolution and represent the process adequately were significantly reduced. We also showed why this method is preferred over a straightforward retrieval of output density matrices from averaged wave-function components. The Monte-Carlo wave-function method provides the further insight that, in the case of little dissipation, a single ```no-jump'' trajectory is enough to find an upper bound on trace distance between the ideal $\chi$ and the simulated one. Such upper bound estimates may be helpful in estimating optimal parameters for experiments. Furthermore, our analysis showed that the mathematical inversion problem occurring in standard process characterisation can be solved in the much lower dimensional vector space. Although identified by our state vector formalism it may be applied to density matrices in usual process tomography.

Our method was demonstrated on the Rydberg mediated two-qubit C-PHASE gate. In future work we plan to address larger quantum systems for which reliable error estimates are needed. Multi-bit gates and the application of consecutive gates together with realistic simulations of error correction codes constitute appealing applications of our numerically efficient method. 

\appendix*
\section{}
Let $\{|j\rangle\rangle :j=1,\dots,D^2\}$ be a linearly independent basis for the $D^2$-dimensional vectors on the principal-ancilla space $P\otimes A$. The initial bipartite system-accilla state vector $|\Psi\rangle\rangle=\sum_{r=1}^D|r\rangle\otimes|r\rangle$ evolves during the Monte Carlo simulation of the process $\mathcal{E}$ into a (stochastic) state vector that can be expanded in this basis
\begin{equation}
|\Psi_{\mathcal{E}}\rangle\rangle=\sum_j\lambda_j|j\rangle\rangle\, .
\end{equation}
Averaging the outer product of many of these states yields the outcome of the master equation evolution of the initial entangled density operator $\mathcal{O}=|\Psi\rangle\rangle\langle\langle\Psi|$. Since $\mathcal{O}=\sum_{r,s=1}^D \mathcal{O}_{rs}\otimes \mathcal{O}_{rs}$ (Eq.~(\ref{eq:superop})) we obtain
\begin{align}
\label{eq:outer1}
\overline{|\Psi_{\mathcal{E}}\rangle\rangle\langle\langle\Psi_{\mathcal{E}}|}&=(\mathcal{E}\otimes \mathcal{I})(\mathcal{O})\nonumber\\
&=\sum_{\rm rs}\mathcal{E}(\mathcal{O}_{\rm rs})\otimes\mathcal{O}_{\rm rs}\nonumber\\
&=\sum_{\rm rs}\sum_{\rm pq}\Lambda_{\rm rs,pq}\mathcal{O}_{\rm pq}\otimes\mathcal{O}_{\rm rs}
\end{align}
where $\{\mathcal{O}_{\rm pq}=|p\rangle\langle q|:p,q=1,\dots,D\}$ is a linearly independent basis for the space of $D\times D$ linear operators. 

In Eq.~(\ref{eq:kappa}) we define 
\begin{equation}
\label{eq:kappa2}
(E_m\otimes I)|\Psi\rangle\rangle=\sum_j \kappa_j^m |j\rangle\rangle\, ,
\end{equation}
and expansion of the outer product of two such states yields
\begin{align}
\label{eq:outer2}
(E_m\otimes I)|\Psi\rangle\rangle\langle\langle\Psi|(E_n^{\dag}\otimes I)&=E_m|r\rangle\langle s|E_n^{\dag}\otimes |r\rangle\langle s|\nonumber\\
&=\sum_{pq}\mathcal{K}_{rs,pq}^{mn}\mathcal{O}_{pq}\otimes\mathcal{O}_{\rm rs},
\end{align}
where we have used the notation defined in Eq.~(\ref{eq:beta}).

In the text, we define $\zeta$ as the solution to Eq.~(\ref{eq:linearzeta}). By forming the outer product of 
\begin{equation}
\label{eq:implyzeta}
\sum_j (\sum_m\kappa_j^m\zeta_m) |j\rangle\rangle=\sum_j \lambda_j |j\rangle\rangle, 
\end{equation}
and averaging over many simulation outcomes, we obtain by Eqs.~(\ref{eq:outer1}) and~(\ref{eq:outer2})
\begin{equation}
\label{eq:proveclaim}
\sum_{mn}\mathcal{K}_{rs,pq}^{mn}\overline{\zeta_m\zeta_n^{\ast}}=\Lambda_{rs,pq}
\end{equation}
for all $r$,$s$. 
 Before Eq.~(7) in the main text we identified the process matrix $\chi$ as the solution to the same equation
$\sum_{mn}\mathcal{K}_{rs,pq}^{mn}\chi_{mn}=\Lambda_{rs,pq}$, and we have thus shown that $\chi$ is directly obtained from the $\zeta$-vectors, $\chi_{\rm mn}=\overline{\zeta_m\zeta_n^{\ast}}$.

\begin{acknowledgments}
The authors gratefully acknowledge discussion with Mark Saffman. This work was supported by the IARPA MQCO program through ARO contract W911NF-10-1-0347.
\end{acknowledgments}


\begin{thebibliography}{23}%
\makeatletter
\providecommand \@ifxundefined [1]{%
 \@ifx{#1\undefined}
}%
\providecommand \@ifnum [1]{%
 \ifnum #1\expandafter \@firstoftwo
 \else \expandafter \@secondoftwo
 \fi
}%
\providecommand \@ifx [1]{%
 \ifx #1\expandafter \@firstoftwo
 \else \expandafter \@secondoftwo
 \fi
}%
\providecommand \natexlab [1]{#1}%
\providecommand \enquote  [1]{``#1''}%
\providecommand \bibnamefont  [1]{#1}%
\providecommand \bibfnamefont [1]{#1}%
\providecommand \citenamefont [1]{#1}%
\providecommand \href@noop [0]{\@secondoftwo}%
\providecommand \href [0]{\begingroup \@sanitize@url \@href}%
\providecommand \@href[1]{\@@startlink{#1}\@@href}%
\providecommand \@@href[1]{\endgroup#1\@@endlink}%
\providecommand \@sanitize@url [0]{\catcode `\\12\catcode `\$12\catcode
  `\&12\catcode `\#12\catcode `\^12\catcode `\_12\catcode `\%12\relax}%
\providecommand \@@startlink[1]{}%
\providecommand \@@endlink[0]{}%
\providecommand \url  [0]{\begingroup\@sanitize@url \@url }%
\providecommand \@url [1]{\endgroup\@href {#1}{\urlprefix }}%
\providecommand \urlprefix  [0]{URL }%
\providecommand \Eprint [0]{\href }%
\providecommand \doibase [0]{http://dx.doi.org/}%
\providecommand \selectlanguage [0]{\@gobble}%
\providecommand \bibinfo  [0]{\@secondoftwo}%
\providecommand \bibfield  [0]{\@secondoftwo}%
\providecommand \translation [1]{[#1]}%
\providecommand \BibitemOpen [0]{}%
\providecommand \bibitemStop [0]{}%
\providecommand \bibitemNoStop [0]{.\EOS\space}%
\providecommand \EOS [0]{\spacefactor3000\relax}%
\providecommand \BibitemShut  [1]{\csname bibitem#1\endcsname}%
\let\auto@bib@innerbib\@empty
\bibitem [{\citenamefont {Nielsen}\ and\ \citenamefont
  {Chuang}(2010)}]{nielsen2010quantum}%
  \BibitemOpen
  \bibfield  {author} {\bibinfo {author} {\bibfnamefont {M.}~\bibnamefont
  {Nielsen}}\ and\ \bibinfo {author} {\bibfnamefont {I.}~\bibnamefont
  {Chuang}},\ }\href@noop {} {\emph {\bibinfo {title} {Quantum computation and
  quantum information}}}\ (\bibinfo  {publisher} {Cambridge university press},\
  \bibinfo {year} {2010})\BibitemShut {NoStop}%
\bibitem [{\citenamefont {Chuang}\ and\ \citenamefont
  {Nielsen}(1997)}]{chaung97proctom}%
  \BibitemOpen
  \bibfield  {author} {\bibinfo {author} {\bibfnamefont {I.~L.}\ \bibnamefont
  {Chuang}}\ and\ \bibinfo {author} {\bibfnamefont {M.~A.}\ \bibnamefont
  {Nielsen}},\ }\href {\doibase 10.1080/09500349708231894} {\bibfield
  {journal} {\bibinfo  {journal} {Journal of Modern Optics}\ }\textbf {\bibinfo
  {volume} {44}},\ \bibinfo {pages} {2455} (\bibinfo {year}
  {1997})}\BibitemShut {NoStop}%
\bibitem [{\citenamefont {Poyatos}\ \emph {et~al.}(1997)\citenamefont
  {Poyatos}, \citenamefont {Cirac},\ and\ \citenamefont
  {Zoller}}]{PhysRevLett.78.390}%
  \BibitemOpen
  \bibfield  {author} {\bibinfo {author} {\bibfnamefont {J.~F.}\ \bibnamefont
  {Poyatos}}, \bibinfo {author} {\bibfnamefont {J.~I.}\ \bibnamefont {Cirac}},
  \ and\ \bibinfo {author} {\bibfnamefont {P.}~\bibnamefont {Zoller}},\ }\href
  {\doibase 10.1103/PhysRevLett.78.390} {\bibfield  {journal} {\bibinfo
  {journal} {Phys. Rev. Lett.}\ }\textbf {\bibinfo {volume} {78}},\ \bibinfo
  {pages} {390} (\bibinfo {year} {1997})}\BibitemShut {NoStop}%
\bibitem [{\citenamefont {Leung}(2003)}]{choiproof_leung}%
  \BibitemOpen
  \bibfield  {author} {\bibinfo {author} {\bibfnamefont {D.~W.}\ \bibnamefont
  {Leung}},\ }\href
  {http://search.ebscohost.com/login.aspx?direct=true&db=afh&AN=8944823&site=ehost-live}
  {\bibfield  {journal} {\bibinfo  {journal} {Journal of Mathematical Physics}\
  }\textbf {\bibinfo {volume} {44}},\ \bibinfo {pages} {528} (\bibinfo {year}
  {2003})}\BibitemShut {NoStop}%
\bibitem [{\citenamefont {D'Ariano}\ and\ \citenamefont
  {Lo~Presti}(2001)}]{PhysRevLett.86.4195}%
  \BibitemOpen
  \bibfield  {author} {\bibinfo {author} {\bibfnamefont {G.~M.}\ \bibnamefont
  {D'Ariano}}\ and\ \bibinfo {author} {\bibfnamefont {P.}~\bibnamefont
  {Lo~Presti}},\ }\href {\doibase 10.1103/PhysRevLett.86.4195} {\bibfield
  {journal} {\bibinfo  {journal} {Phys. Rev. Lett.}\ }\textbf {\bibinfo
  {volume} {86}},\ \bibinfo {pages} {4195} (\bibinfo {year}
  {2001})}\BibitemShut {NoStop}%
\bibitem [{\citenamefont {Altepeter}\ \emph {et~al.}(2003)\citenamefont
  {Altepeter}, \citenamefont {Branning}, \citenamefont {Jeffrey}, \citenamefont
  {Wei}, \citenamefont {Kwiat}, \citenamefont {Thew}, \citenamefont {O'Brien},
  \citenamefont {Nielsen},\ and\ \citenamefont
  {White}}]{PhysRevLett.90.193601}%
  \BibitemOpen
  \bibfield  {author} {\bibinfo {author} {\bibfnamefont {J.~B.}\ \bibnamefont
  {Altepeter}}, \bibinfo {author} {\bibfnamefont {D.}~\bibnamefont {Branning}},
  \bibinfo {author} {\bibfnamefont {E.}~\bibnamefont {Jeffrey}}, \bibinfo
  {author} {\bibfnamefont {T.~C.}\ \bibnamefont {Wei}}, \bibinfo {author}
  {\bibfnamefont {P.~G.}\ \bibnamefont {Kwiat}}, \bibinfo {author}
  {\bibfnamefont {R.~T.}\ \bibnamefont {Thew}}, \bibinfo {author}
  {\bibfnamefont {J.~L.}\ \bibnamefont {O'Brien}}, \bibinfo {author}
  {\bibfnamefont {M.~A.}\ \bibnamefont {Nielsen}}, \ and\ \bibinfo {author}
  {\bibfnamefont {A.~G.}\ \bibnamefont {White}},\ }\href {\doibase
  10.1103/PhysRevLett.90.193601} {\bibfield  {journal} {\bibinfo  {journal}
  {Phys. Rev. Lett.}\ }\textbf {\bibinfo {volume} {90}},\ \bibinfo {pages}
  {193601} (\bibinfo {year} {2003})}\BibitemShut {NoStop}%
\bibitem [{\citenamefont {D'Ariano}\ and\ \citenamefont
  {Lo~Presti}(2003)}]{PhysRevLett.91.047902}%
  \BibitemOpen
  \bibfield  {author} {\bibinfo {author} {\bibfnamefont {G.~M.}\ \bibnamefont
  {D'Ariano}}\ and\ \bibinfo {author} {\bibfnamefont {P.}~\bibnamefont
  {Lo~Presti}},\ }\href {\doibase 10.1103/PhysRevLett.91.047902} {\bibfield
  {journal} {\bibinfo  {journal} {Phys. Rev. Lett.}\ }\textbf {\bibinfo
  {volume} {91}},\ \bibinfo {pages} {047902} (\bibinfo {year}
  {2003})}\BibitemShut {NoStop}%
\bibitem [{\citenamefont {Mohseni}\ and\ \citenamefont
  {Lidar}(2006)}]{PhysRevLett.97.170501}%
  \BibitemOpen
  \bibfield  {author} {\bibinfo {author} {\bibfnamefont {M.}~\bibnamefont
  {Mohseni}}\ and\ \bibinfo {author} {\bibfnamefont {D.~A.}\ \bibnamefont
  {Lidar}},\ }\href {\doibase 10.1103/PhysRevLett.97.170501} {\bibfield
  {journal} {\bibinfo  {journal} {Phys. Rev. Lett.}\ }\textbf {\bibinfo
  {volume} {97}},\ \bibinfo {pages} {170501} (\bibinfo {year}
  {2006})}\BibitemShut {NoStop}%
\bibitem [{\citenamefont {Mohseni}\ and\ \citenamefont
  {Lidar}(2007)}]{PhysRevA.75.062331}%
  \BibitemOpen
  \bibfield  {author} {\bibinfo {author} {\bibfnamefont {M.}~\bibnamefont
  {Mohseni}}\ and\ \bibinfo {author} {\bibfnamefont {D.~A.}\ \bibnamefont
  {Lidar}},\ }\href {\doibase 10.1103/PhysRevA.75.062331} {\bibfield  {journal}
  {\bibinfo  {journal} {Phys. Rev. A}\ }\textbf {\bibinfo {volume} {75}},\
  \bibinfo {pages} {062331} (\bibinfo {year} {2007})}\BibitemShut {NoStop}%
\bibitem [{\citenamefont {Mohseni}\ \emph {et~al.}(2008)\citenamefont
  {Mohseni}, \citenamefont {Rezakhani},\ and\ \citenamefont
  {Lidar}}]{PhysRevA.77.032322}%
  \BibitemOpen
  \bibfield  {author} {\bibinfo {author} {\bibfnamefont {M.}~\bibnamefont
  {Mohseni}}, \bibinfo {author} {\bibfnamefont {A.~T.}\ \bibnamefont
  {Rezakhani}}, \ and\ \bibinfo {author} {\bibfnamefont {D.~A.}\ \bibnamefont
  {Lidar}},\ }\href {\doibase 10.1103/PhysRevA.77.032322} {\bibfield  {journal}
  {\bibinfo  {journal} {Phys. Rev. A}\ }\textbf {\bibinfo {volume} {77}},\
  \bibinfo {pages} {032322} (\bibinfo {year} {2008})}\BibitemShut {NoStop}%
\bibitem [{\citenamefont {Carmichael}(1993)}]{carmichael1993open}%
  \BibitemOpen
  \bibfield  {author} {\bibinfo {author} {\bibfnamefont {H.}~\bibnamefont
  {Carmichael}},\ }\href@noop {} {\emph {\bibinfo {title} {An open systems
  approach to Quantum Optics}}},\ Vol.~\bibinfo {volume} {18}\ (\bibinfo
  {publisher} {Springer},\ \bibinfo {year} {1993})\BibitemShut {NoStop}%
\bibitem [{\citenamefont {Dalibard}\ \emph {et~al.}(1992)\citenamefont
  {Dalibard}, \citenamefont {Castin},\ and\ \citenamefont
  {M\o{}lmer}}]{PhysRevLett.68.580}%
  \BibitemOpen
  \bibfield  {author} {\bibinfo {author} {\bibfnamefont {J.}~\bibnamefont
  {Dalibard}}, \bibinfo {author} {\bibfnamefont {Y.}~\bibnamefont {Castin}}, \
  and\ \bibinfo {author} {\bibfnamefont {K.}~\bibnamefont {M\o{}lmer}},\ }\href
  {\doibase 10.1103/PhysRevLett.68.580} {\bibfield  {journal} {\bibinfo
  {journal} {Phys. Rev. Lett.}\ }\textbf {\bibinfo {volume} {68}},\ \bibinfo
  {pages} {580} (\bibinfo {year} {1992})}\BibitemShut {NoStop}%
\bibitem [{\citenamefont {M{\o}lmer}\ \emph {et~al.}(1993)\citenamefont
  {M{\o}lmer}, \citenamefont {Castin},\ and\ \citenamefont
  {Dalibard}}]{Molmer:93}%
  \BibitemOpen
  \bibfield  {author} {\bibinfo {author} {\bibfnamefont {K.}~\bibnamefont
  {M{\o}lmer}}, \bibinfo {author} {\bibfnamefont {Y.}~\bibnamefont {Castin}}, \
  and\ \bibinfo {author} {\bibfnamefont {J.}~\bibnamefont {Dalibard}},\ }\href
  {\doibase 10.1364/JOSAB.10.000524} {\bibfield  {journal} {\bibinfo  {journal}
  {J. Opt. Soc. Am. B}\ }\textbf {\bibinfo {volume} {10}},\ \bibinfo {pages}
  {524} (\bibinfo {year} {1993})}\BibitemShut {NoStop}%
\bibitem [{\citenamefont {Dum}\ \emph {et~al.}(1992)\citenamefont {Dum},
  \citenamefont {Zoller},\ and\ \citenamefont {Ritsch}}]{PhysRevA.45.4879}%
  \BibitemOpen
  \bibfield  {author} {\bibinfo {author} {\bibfnamefont {R.}~\bibnamefont
  {Dum}}, \bibinfo {author} {\bibfnamefont {P.}~\bibnamefont {Zoller}}, \ and\
  \bibinfo {author} {\bibfnamefont {H.}~\bibnamefont {Ritsch}},\ }\href
  {\doibase 10.1103/PhysRevA.45.4879} {\bibfield  {journal} {\bibinfo
  {journal} {Phys. Rev. A}\ }\textbf {\bibinfo {volume} {45}},\ \bibinfo
  {pages} {4879} (\bibinfo {year} {1992})}\BibitemShut {NoStop}%
\bibitem [{\citenamefont {Gardiner}\ and\ \citenamefont
  {Zoller}(2004)}]{gardiner2004quantum}%
  \BibitemOpen
  \bibfield  {author} {\bibinfo {author} {\bibfnamefont {C.}~\bibnamefont
  {Gardiner}}\ and\ \bibinfo {author} {\bibfnamefont {P.}~\bibnamefont
  {Zoller}},\ }\href@noop {} {\emph {\bibinfo {title} {Quantum noise: a
  handbook of Markovian and non-Markovian quantum stochastic methods with
  applications to quantum optics}}},\ Vol.~\bibinfo {volume} {56}\ (\bibinfo
  {publisher} {Springer},\ \bibinfo {year} {2004})\BibitemShut {NoStop}%
\bibitem [{\citenamefont {Kraus}\ \emph {et~al.}(1983)\citenamefont {Kraus},
  \citenamefont {B{\"o}hm}, \citenamefont {Dollard},\ and\ \citenamefont
  {Wootters}}]{kraus1983states}%
  \BibitemOpen
  \bibfield  {author} {\bibinfo {author} {\bibfnamefont {K.}~\bibnamefont
  {Kraus}}, \bibinfo {author} {\bibfnamefont {A.}~\bibnamefont {B{\"o}hm}},
  \bibinfo {author} {\bibfnamefont {J.}~\bibnamefont {Dollard}}, \ and\
  \bibinfo {author} {\bibfnamefont {W.}~\bibnamefont {Wootters}},\ }in\
  \href@noop {} {\emph {\bibinfo {booktitle} {States, Effects, and Operations
  Fundamental Notions of Quantum Theory}}},\ Vol.\ \bibinfo {volume} {190}\
  (\bibinfo {year} {1983})\BibitemShut {NoStop}%
\bibitem [{Note1()}]{Note1}%
  \BibitemOpen
  \bibinfo {note} {In a multi-qubit system, an appropriate operator basis might
  be tensor products of the identity and Pauli operators for each
  qubit.}\BibitemShut {Stop}%
\bibitem [{\citenamefont {Bardou}\ \emph {et~al.}(1994)\citenamefont {Bardou},
  \citenamefont {Bouchaud}, \citenamefont {Emile}, \citenamefont {Aspect},\
  and\ \citenamefont {Cohen-Tannoudji}}]{PhysRevLett.72.203}%
  \BibitemOpen
  \bibfield  {author} {\bibinfo {author} {\bibfnamefont {F.}~\bibnamefont
  {Bardou}}, \bibinfo {author} {\bibfnamefont {J.~P.}\ \bibnamefont
  {Bouchaud}}, \bibinfo {author} {\bibfnamefont {O.}~\bibnamefont {Emile}},
  \bibinfo {author} {\bibfnamefont {A.}~\bibnamefont {Aspect}}, \ and\ \bibinfo
  {author} {\bibfnamefont {C.}~\bibnamefont {Cohen-Tannoudji}},\ }\href
  {\doibase 10.1103/PhysRevLett.72.203} {\bibfield  {journal} {\bibinfo
  {journal} {Phys. Rev. Lett.}\ }\textbf {\bibinfo {volume} {72}},\ \bibinfo
  {pages} {203} (\bibinfo {year} {1994})}\BibitemShut {NoStop}%
\bibitem [{\citenamefont {Gilchrist}\ \emph {et~al.}(2005)\citenamefont
  {Gilchrist}, \citenamefont {Langford},\ and\ \citenamefont
  {Nielsen}}]{PhysRevA.71.062310}%
  \BibitemOpen
  \bibfield  {author} {\bibinfo {author} {\bibfnamefont {A.}~\bibnamefont
  {Gilchrist}}, \bibinfo {author} {\bibfnamefont {N.~K.}\ \bibnamefont
  {Langford}}, \ and\ \bibinfo {author} {\bibfnamefont {M.~A.}\ \bibnamefont
  {Nielsen}},\ }\href {\doibase 10.1103/PhysRevA.71.062310} {\bibfield
  {journal} {\bibinfo  {journal} {Phys. Rev. A}\ }\textbf {\bibinfo {volume}
  {71}},\ \bibinfo {pages} {062310} (\bibinfo {year} {2005})}\BibitemShut
  {NoStop}%
\bibitem [{\citenamefont {Reiter}\ and\ \citenamefont
  {S\o{}rensen}(2012)}]{PhysRevA.85.032111}%
  \BibitemOpen
  \bibfield  {author} {\bibinfo {author} {\bibfnamefont {F.}~\bibnamefont
  {Reiter}}\ and\ \bibinfo {author} {\bibfnamefont {A.~S.}\ \bibnamefont
  {S\o{}rensen}},\ }\href {\doibase 10.1103/PhysRevA.85.032111} {\bibfield
  {journal} {\bibinfo  {journal} {Phys. Rev. A}\ }\textbf {\bibinfo {volume}
  {85}},\ \bibinfo {pages} {032111} (\bibinfo {year} {2012})}\BibitemShut
  {NoStop}%
\bibitem [{\citenamefont {Johnson}\ \emph {et~al.}(2008)\citenamefont
  {Johnson}, \citenamefont {Urban}, \citenamefont {Henage}, \citenamefont
  {Isenhower}, \citenamefont {Yavuz}, \citenamefont {Walker},\ and\
  \citenamefont {Saffman}}]{PhysRevLett.100.113003}%
  \BibitemOpen
  \bibfield  {author} {\bibinfo {author} {\bibfnamefont {T.~A.}\ \bibnamefont
  {Johnson}}, \bibinfo {author} {\bibfnamefont {E.}~\bibnamefont {Urban}},
  \bibinfo {author} {\bibfnamefont {T.}~\bibnamefont {Henage}}, \bibinfo
  {author} {\bibfnamefont {L.}~\bibnamefont {Isenhower}}, \bibinfo {author}
  {\bibfnamefont {D.~D.}\ \bibnamefont {Yavuz}}, \bibinfo {author}
  {\bibfnamefont {T.~G.}\ \bibnamefont {Walker}}, \ and\ \bibinfo {author}
  {\bibfnamefont {M.}~\bibnamefont {Saffman}},\ }\href {\doibase
  10.1103/PhysRevLett.100.113003} {\bibfield  {journal} {\bibinfo  {journal}
  {Phys. Rev. Lett.}\ }\textbf {\bibinfo {volume} {100}},\ \bibinfo {pages}
  {113003} (\bibinfo {year} {2008})}\BibitemShut {NoStop}%
\bibitem [{\citenamefont {Saffman}\ \emph {et~al.}(2011)\citenamefont
  {Saffman}, \citenamefont {Zhang}, \citenamefont {Gill}, \citenamefont
  {Isenhower},\ and\ \citenamefont {Walker}}]{saffman2011rydberg}%
  \BibitemOpen
  \bibfield  {author} {\bibinfo {author} {\bibfnamefont {M.}~\bibnamefont
  {Saffman}}, \bibinfo {author} {\bibfnamefont {X.}~\bibnamefont {Zhang}},
  \bibinfo {author} {\bibfnamefont {A.}~\bibnamefont {Gill}}, \bibinfo {author}
  {\bibfnamefont {L.}~\bibnamefont {Isenhower}}, \ and\ \bibinfo {author}
  {\bibfnamefont {T.}~\bibnamefont {Walker}},\ }in\ \href@noop {} {\emph
  {\bibinfo {booktitle} {Journal of Physics: Conference Series}}},\ Vol.\
  \bibinfo {volume} {264}\ (\bibinfo {organization} {IOP Publishing},\ \bibinfo
  {year} {2011})\ p.\ \bibinfo {pages} {012023}\BibitemShut {NoStop}%
\bibitem [{\citenamefont {Castin}\ and\ \citenamefont
  {M{\o}lmer}(1996)}]{PhysRevA.54.5275}%
  \BibitemOpen
  \bibfield  {author} {\bibinfo {author} {\bibfnamefont {Y.}~\bibnamefont
  {Castin}}\ and\ \bibinfo {author} {\bibfnamefont {K.}~\bibnamefont
  {M{\o}lmer}},\ }\href {\doibase 10.1103/PhysRevA.54.5275} {\bibfield
  {journal} {\bibinfo  {journal} {Phys. Rev. A}\ }\textbf {\bibinfo {volume}
  {54}},\ \bibinfo {pages} {5275} (\bibinfo {year} {1996})}\BibitemShut
  {NoStop}%
\end{thebibliography}
\end{document}